\documentclass[dvips]{acta}
\usepackage{supertabular,lscape,epsfig}
\usepackage{amssymb}
\usepackage{amsmath}
\usepackage[latin2]{inputenc}

\SetPages{0}{0}

\SetVol{0}{2016}

\begin{document}

\begin{Titlepage}
\Title{New transit observations for HAT-P-30~b, HAT-P-37~b, TrES-5~b, WASP-28~b, WASP-36~b, and WASP-39~b}
\Author{G.~~M~a~c~i~e~j~e~w~s~k~i$^{1}$, ~ D.~~D~i~m~i~t~r~o~v$^{2}$, ~ L.~~M~a~n~c~i~n~i$^{3,4}$, ~ J.~~S~o~u~t~h~w~o~r~t~h$^{5}$, ~ S.~~C~i~c~e~r~i$^{3}$, ~ G.~~D'~A~g~o$^{6}$, ~ I.~~B~r~u~n~i$^{7}$, ~ St.~~R~a~e~t~z$^{8,9}$, ~ G.~~N~o~w~a~k$^{1,10,11}$, ~ J.~~O~h~l~e~r~t$^{12,13}$, ~ D.~~P~u~c~h~a~l~s~k~i$^{1}$, ~ G.~~S~a~r~a~l$^{14,15}$, ~ E.~~D~e~r~m~a~n$^{16}$, ~ R.~~P~e~t~r~u~c~c~i\footnote{Visiting Astronomer, Complejo Astron\'omico El Leoncito operated under agreement between the Consejo Nacional de Investigaciones Cient\'ificas y T\'ecnicas de la Rep\'ublica Argentina and the National Universities of La Plata, C\'ordoba and San Juan.}$^{17,18}$, ~ E.~~J~o~f~r~\'e$^{17,18}$, M.~~S~e~e~l~i~g~e~r$^{8}$, ~ and ~ T.~~H~e~n~n~i~n~g$^{3}$}
{$^1$Centre for Astronomy, Faculty of Physics, Astronomy and Informatics,
         Nicolaus Copernicus University, Grudziadzka 5, 87-100 Torun, Poland,
         e-mail: gmac@umk.pl\\
 $^2$Institute of Astronomy, Bulgarian Academy of Sciences, 72 Tsarigradsko Chausse Blvd., 1784 Sofia, Bulgaria\\
 $^3$Max Planck Institute for Astronomy, K\"onigstuhl 17, 69117 Heidelberg, Germany\\
 $^4$INAF - Osservatorio Astrofisico di Torino, via Osservatorio 20, 10025 - Pino Torinese, Italy\\
 $^5$Astrophysics Group, Keele University, Staffordshire, ST5 5BG, UK\\
 $^6$Istituto Internazionale per gli Alti Studi Scientifici (IIASS), 84019 Vietri Sul Mare, Italy\\
 $^7$INAF - Osservatorio Astronomico di Bologna, via Ranzani 1, 40127 Bologna, Italy\\
 $^8$Astrophysikalisches Institut und Universit\"ats-Sternwarte, Schillerg\"asschen 2--3, 07745 Jena, Germany\\
 $^{9}$European Space Agency, ESTEC, SRE-S, Keplerlaan 1, 2201 AZ Noordwijk, The Netherlands\\
 $^{10}$Instituto de Astrof\'isica de Canarias, C/ v\'ia L\'actea, s/n, E-38205 La Laguna, Tenerife, Spain\\
 $^{11}$Departamento de Astrof\'isica, Universidad de La Laguna, Av. Astrof\'isico Francisco S\'anchez, s/n, E-38206 La Laguna, Tenerife, Spain\\
 $^{12}$Michael Adrian Observatorium, Astronomie Stiftung Trebur, 65468 Trebur, Germany\\
 $^{13}$University of Applied Sciences, Technische Hochschule Mittelhessen, 61169 Friedberg, Germany\\
 $^{14}$University of Geneva, Department of Astronomy, Ch. d'Ecogia 16, 1290 Versoix, Switzerland\\
 $^{15}$Istanbul University, Graduate School of Science and Engineering, Bozdogan Kemeri Cad. 8, Vezneciler-Istanbul, Turkey\\
 $^{16}$Astronomy and Space Sciences Department, Ankara University, 06100 Tando\v{g}an, Ankara, Turkey\\
 $^{17}$Observatorio Astron\'omico de C\'ordoba (OAC), Laprida 854, X5000 BGR C\'ordoba, Argentina\\
 $^{18}$Consejo Nacional de Investigaciones Cient\'ificas y T\'ecnicas (CONICET), Argentina}

\Received{2016, submitted to Acta Astronomica}
\end{Titlepage}

\Abstract{We present new transit light curves for planets in six extrasolar planetary systems. They were acquired with 0.4-2.2 m telescopes located in west Asia, Europe, and South America. When combined with literature data, they allowed us to redetermine system parameters in a homogeneous way. Our results for individual systems are in agreement with values reported in previous studies. We refined transit ephemerides and reduced uncertainties of orbital periods by a factor between 2 and 7. No sign of any variations in transit times was detected for the planets studied.}{planetary systems -- planets and satellites: individual: HAT-P-30 b, HAT-P-37 b, TrES-5 b, WASP-28 b, WASP-36 b, WASP-39 b}


\section{Introduction}

If the orbital plane of an exoplanet is orientated almost perpendicular to the plane of the sky, the planet is likely to transit its host star which is manifested as small regular dips in the observed stellar flux. The transit technique has become an efficient tool for confirming the planetary nature of planet candidates discovered with the radial velocity (RV) method (Charbonneau \etal 2000, Henry \etal 2000) and for discovering new exoplanets (Konacki \etal 2003). For transiting planets, there is a unique opportunity to determine its mass and radius, and hence the mean density, which is a key parameter for studies of planetary internal structure. Modelling can then constrain the chemical composition and the mass of the solid core (\eg Nettelmann \etal 2010). The Rossiter-McLaughlin effect can provide information about the spin axis and orbital plane alignment and, hence, the dynamics in the system (\eg Queloz \etal 2000). One can also study the atmospheric composition with transmission spectroscopy (\eg Charbonneau \etal 2002) and indirectly determine the dayside emission from a planet by detecting the occultation (\eg Deming \etal 2005). High-precision follow-up photometric observations of transits are used to refine system parameters, better knowledge of which may be beneficial for statistical studies. Long-term transit timing may reveal a departure from a linear ephemeris due to an orbital decay caused by planet-star tidal interactions (\eg Murgas \etal 2014) or due to gravitational perturbations induced by an unseen planetary companion (\eg Ballard \etal 2011). In this paper, we present new photometric time series for transits of six exoplanets: HAT-P-30~b, HAT-P-37~b, TrES-5~b, WASP-28~b, WASP-36~b, and WASP-39~b. For half of them, our data are the first follow-up observations acquired a number of orbital periods (epochs) after their first studies.

HAT-P-30~b (also known as WASP-51~b) was independently discovered by both the Hungarian-made Automatic Telescope Network (HAT, Johnson \etal 2011) and the Wide Angle Search for Planets (WASP) project (Enoch \etal 2011). According to Johnson \etal (2011), it has a radius of $1.340\pm0.065$ $R_{Jup}$ (Jupiter radius) and a mass of $0.711\pm0.028$ $M_{Jup}$ (Jupiter mass) that gives a relatively low mean density of $0.279 \pm 0.038 \, \rho_{Jup}$ (Jupiter density). The orbital period is 2.81 days. The host star BD\,+06~1909 ($V=10.35$ mag) is a 1 Gyr-old dwarf of spectral type F (Johnson \etal 2011). No follow-up studies have been published for this system so far.

HAT-P-37~b orbits the G-type dwarf GSC 03553-00723 ($V=13.23$ mag) with a period of 2.80 days (Bakos \etal 2012). According to the discovery paper, it has a mass of $1.17\pm0.10$ $M_{Jup}$, a radius of $1.18\pm0.08$ $R_{Jup}$, and a mean density of $0.67 \pm 0.14 \, \rho_{Jup}$. To date, the system has not been the subject of any follow-up observations.

The TrES-5 system is composed of the cool G/K dwarf GSC 03949-00967 ($V=13.72$ mag), one of the faintest transiting planet host stars discovered from the ground, and a typical hot Jupiter on a 1.48-day circular orbit (Mandushev  \etal 2011). The planet's mass and radius were found to be $1.778 \pm 0.063$ $M_{Jup}$ and $1.209 \pm 0.021$ $R_{Jup}$, respectively. Those quantities result in a mean planetary density of $0.94 \pm 0.06 \, \rho_{Jup}$. Mislis \etal (2015) presented the first photometric follow-up observations that confirm the values of the system parameters published in the discovery paper.

WASP-28 b is a classical bloated Jupiter-mass planet. It has a mass of $0.907 \pm 0.043$ $M_{Jup}$, a radius of $1.213 \pm 0.042$ $R_{Jup}$, and a mean density of $0.508 \pm 0.047$ $\rho_{Jup}$ (Anderson \etal 2015). The host star 2MASS J23342787-0134482 ($V=12.03$ mag) is a dwarf of spectral type F8. The planetary orbit, with a period of 3.41 days, was found to be prograde and aligned (Anderson \etal 2015). System parameters were redetermined by Petrucci \etal (2015) who analyzed a set of 4 new light curves, enhanced with 11 amateur time series.

WASP-36 b was found to be a planet with a mass of $2.30\pm0.07$ $M_{Jup}$, a radius of $1.28\pm0.03$ $R_{Jup}$, and a mean density of $1.096 \pm 0.067$ $\rho_{Jup}$ (Smith \etal 2012). It is on a 1.54-day orbit around the metal-poor G2 dwarf GSC 05442-00759 ($V=12.7$ mag). The only photometric follow-up observations come from the discovery paper. The system has remained unstudied since then.

With a mass of $0.28\pm0.03$ $M_{Jup}$ and a radius of $1.27\pm0.04$ $R_{Jup}$, WASP-39~b belongs to a population of inflated giant planets (Faedi \etal 2011). Its mean density was reported to be $0.14\pm0.02$ $\rho_{Jup}$. It orbits the G8 main-sequence star GSC 04980-00761 ($V=12.09$ mag) with a period of 4.06 days. Ricci \etal (2015) present the first follow-up photometric time series for transits that were used to redetermine the system parameters.


\section{Observations}

We acquired 15 new transit light curves with nine 0.4--2.2 m telescopes located in Europe, west Asia, and South America. Some observations were performed without any filter in order to increase the throughput of the instrument, and hence to get more precise data. In such cases, the effective maximum of the instrument sensitivity was found to fall in between $V$ and $R$ bands. New light curves are presented in Fig.~1. The summary of the individual time series is given in Table 1. To quantify the quality of a given light curve, we adopted the photometric noise rate ($pnr$) from Fulton \etal (2011) defined as
\begin{equation}
  pnr = \frac{rms}{\sqrt{\Gamma}}
\end{equation}
where $rms$ is the root mean square of the flux residuals from the model light curve, and $\Gamma$ is the median number of exposures per minute. Below we give the detailed description of the individual runs and the instruments used.

\begin{figure}[thb]
\begin{center}
\includegraphics[width=1.0\textwidth]{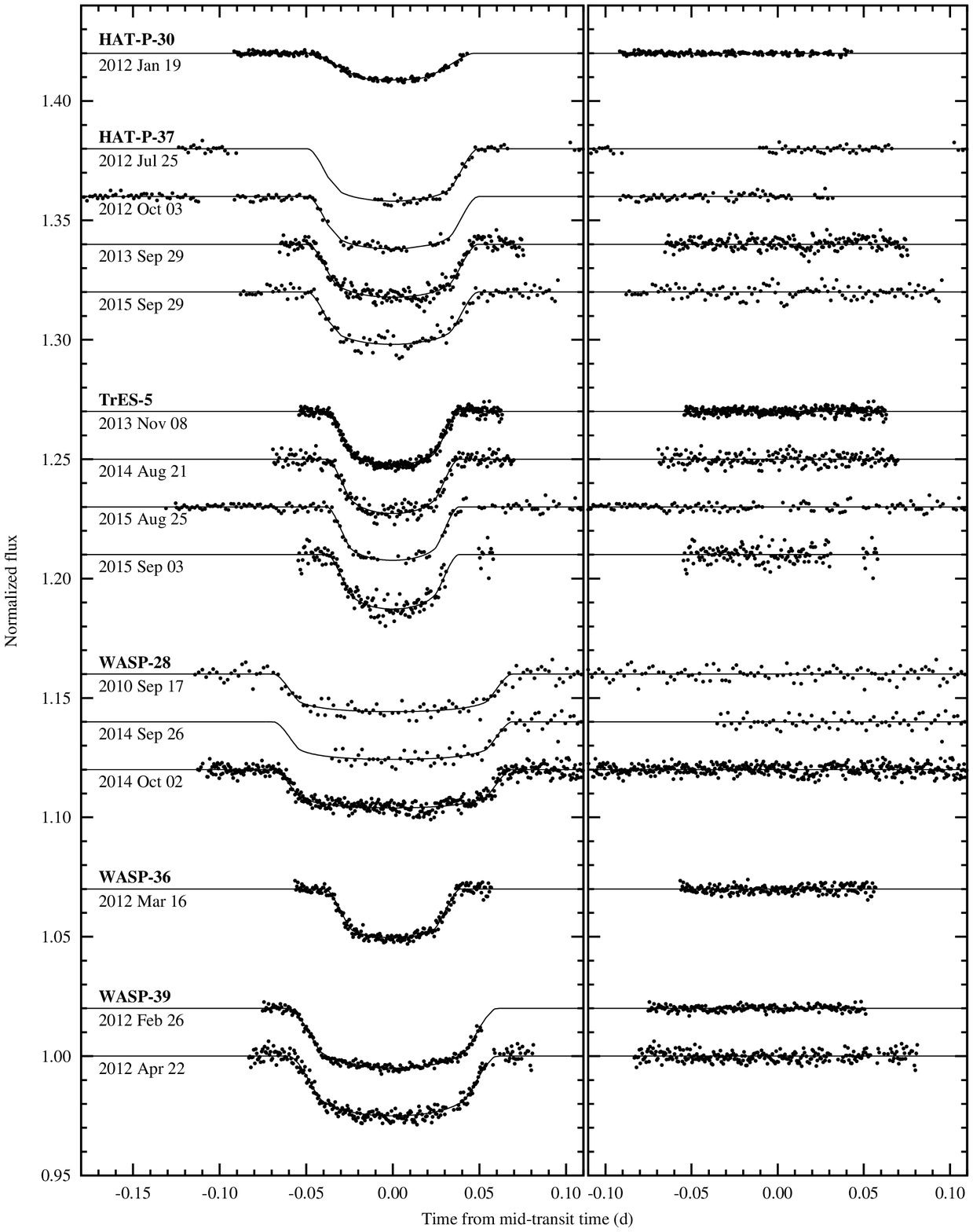}
\end{center}
\FigCap{Left panel: new transit light curves for the systems studied, sorted by object name and observation date. The best-fit models are plotted with continuous lines. Right panel: residuals from the best-fit models. The light curves, acquired for WASP-28 on 2010 September 17 and 2014 September 26, are binned into 3 minute bins for greater clarity of the plot.}
\end{figure}

\MakeTable{l l l c c c c}{12.5cm}{Summary for new transit light curves. Date UT is given for the middle of the transit, $X$ is the airmass change during transit observations, $N_{obs}$ is the number of useful scientific exposures, $\Gamma$ is the median number of exposures per minute, $pnr$ is the photometric scatter in parts per thousand of the normalized flux per minute of observation.}
{\hline
Date UT  & Telescope & Filter  &  $X$       & $N_{obs}$ & $\Gamma$ & $pnr$\\
\hline
\multicolumn{7}{c}{HAT-P-30 b}\\
 2012 Jan 19  & Rozhen 2.0 m  & $R_{\rm{C}}$ & $1.26\rightarrow1.23\rightarrow1.49$ & 180 & 1.22  &  0.55\\
\multicolumn{7}{c}{HAT-P-37 b}\\
 2012 Jul 25  & CAO 1.23 m  & $R_{\rm{C}}$ & $1.11\rightarrow1.30$ & 65 & 0.37  &  1.98\\
 2012 Oct 03  & Cassini 1.52 m  & Gunn-$r$ & $1.01\rightarrow1.74$ & 104 & 0.47  &  1.61\\
 2013 Sep 29  & Piwnice 0.6 m  & $-$ & $1.00\rightarrow1.22$ & 204 &3.99  &  2.00\\
 2015 Sep 29  & OGS 1.0 m  & $R_{\rm{C}}$ & $1.10\rightarrow2.20$ & 108 & 0.41  &  3.68\\
\multicolumn{7}{c}{TrES-5 b}\\
 2013 Nov 08  & Rozhen 2.0 m  & $R_{\rm{C}}$ & $1.21\rightarrow1.75$ & 301 & 1.82 & 0.96\\
 2014 Aug 21  & Piwnice 0.6 m  & $-$ & $1.02\rightarrow1.01\rightarrow1.06$ & 184 & 0.95 & 2.28\\
 2015 Aug 25  & Cassini 1.52 m  & $R_{\rm{C}}$ & $1.05\rightarrow1.04\rightarrow1.46$ & 133 & 1.46 & 1.99\\
 2015 Sep 03  & Piwnice 0.6 m  & $-$ & $1.01\rightarrow1.06$ & 137 & 1.09 & 2.98\\
\multicolumn{7}{c}{WASP-28 b}\\
 2010 Sep 17  & Ankara 0.4 m  & $R_{\rm{C}}$ & $1.80\rightarrow1.33\rightarrow2.54$ & 384 & 3.32 & 4.03\\
 2014 Sep 26  & THG 0.4 m  & $R_{\rm{J}}$ & $1.24\rightarrow1.16\rightarrow1.39$ & 207 & 1.00 & 4.76\\
 2014 Oct 02  & Trebur 1.2 m  & $-$ & $1.83\rightarrow1.60\rightarrow2.66$ & 433 & 1.39 & 1.70\\
\multicolumn{7}{c}{WASP-36 b}\\
 2012 Mar 16  & Rozhen 2.0 m  & $R_{\rm{C}}$ & $1.55\rightarrow2.16$ & 200 & 1.22  &  1.31\\
\multicolumn{7}{c}{WASP-39 b}\\
 2012 Feb 26  & CAO 2.2 m  & $R_{\rm{C}}$ & $1.41\rightarrow1.32\rightarrow1.47$ & 186 & 1.02  &  1.05\\
 2012 Apr 22  & Rozhen 2.0 m  & $R_{\rm{C}}$ & $1.53\rightarrow1.42\rightarrow1.81$ & 258 & 1.22  &  1.80\\
\hline
}

\subsection{HAT-P-30 b}

We acquired a high quality light curve for the transit of HAT-P-30 b on 2012 January 19 with the 2.0 m Ritchey-Chr\'etien-Coud\'e telescope at the National Astronomical Observatory Rozhen (Bulgaria). A Roper Scientific VerArray 1300B CCD camera was used as the detector. It has $1340 \times 1300$ pixels of 20 $\mu$m size. In the focal reducer mode, the original field of view (FoV) has a diameter of about $16'$. To shorten the readout time, the FoV was windowed to $12'.2 \times 12'.2$. The acquisition of the time series began at 22:05 January 18 UT, 65 min before the transit, and ended just after the fourth contact at 01:20 January 19 UT. It was obtained through a Cousins $R$ filter with an exposure time of 30 s. Observing conditions were photometric. The telescope was defocused and stellar profiles had donut-like shapes with a diameter of about $35''$. The telescope pointing was kept fixed within $4''$ thanks to manual guiding.

\subsection{HAT-P-37 b}

We used the 0.6 m Cassegrain telescope at the Centre for Astronomy of the Nicolaus Copernicus University in Piwnice, near Toru\'n (Poland), to observe a complete transit of HAT-P-37~b on 2013 September 29. The instrument was equipped with an SBIG STL-1001 CCD camera, which had $1024 \times 1024$ 24-$\mu$m size pixels and a FoV of $11'.8\times11'.8$. Observations began at 17:37 UT, 30 min before the ingress, and ended at 21:00 UT, 40 min after the egress. The binning mode of $2\times2$ was used for fast readout. The exposure time was 12 s. The time series was acquired without any filter. The instrument was guided by manual compensations of inaccuracies of the tracking system.

A transit on 2015 September 29 was observed with the 1 m ESA's Optical Ground Station (OGS) telescope at the Observatorio del Teide (Spain). A spectrograph in imaging mode, equipped with a Roper CCD matrix with $2048 \times 2048$ 13.5-$\mu$m size pixels, was used for acquisition. The FoV was circular with a diameter of $13.8'$. Collecting scientific data started at 20:12 September 29 UT, about 60 min before the expected beginning of the transit. Observations were ended at 00:36 September 30 UT, 66 min after the event. Exposures were acquired through the Cousins $R$ filter with an exposure time of 60 s. An additional 85 s were used for readout, resulting in a relatively low cadence. The telescope was not guided. During the run, the observed field drifted by $14''$ from the initial position.

We also acquired two incomplete light curves for transits on 2012 July 25 and October 3. The first dataset was obtained with the 1.23 m telescope at the Calar Alto Observatory (CAO, Spain). A DLR-MKIII CCD camera, equipped with a $4096 \times 4112$ pixels matrix, was used as detector. Exposures were windowed to $2296 \times 1812$ pixels, resulting in a $11'.8 \times 9'.4$ FoV. Observations started at 20:50 July 25 UT, and then were interrupted by clouds. Observations were resumed about 15 min before the predicted middle of the transit, and were continued till 02:24 July 26 UT with an additional 50 min long gap after the transit. Exposures of 130-s length were acquired through a Cousins $R$ filter.

The second incomplete light curve was obtained with the 1.52 m Cassini Telescope at the Astronomical Observatory of Bologna in Loiano (Italy). The FoV of the Bologna Faint Object Spectrograph \& Camera (BFOSC) imager, equipped with a $1300 \times 1340$ pixels matrix, was windowed to $9'.4 \times 4'.0$. The photometric time series, which was acquired in the Gunn-$r$ filter with 120 s long individual exposures, started on 2012 October 3 at 18:00 UT, 3 h before the expected first contact. Observations were interrupted by clouds at 23:03 UT, just before the third contact.

\subsection{TrES-5 b}

A millimagnitude-precision light curve was acquired with the 2.0 m telescope at Rozhen on 2013 November 8. We used the original FoV of the camera without the focal reducer, \ie $5'.6 \times 5'.6$. Data acquisition started at 18:41 UT, 26 min before the beginning of the transit, and ended at 21:30 UT, 37 min after the event. Thirty-second exposures were acquired through the Cousins $R$ filter. Observing conditions were photometric. Stellar images drifted by no more than $5''$ thanks to the auto-guiding system.

The light curves from 2014 August 21 and 2015 September 03 were acquired without any filter with the 0.6 m Cassegrain telescope at the Centre for Astronomy of the Nicolaus Copernicus University. During the first run, the observing conditions were stable. The run started at 20:05 UT and ended at 23:25 UT. The host star was monitored for 50 min before and after the transit. The second run started at 19:46 UT, 30 min before the transit and was finally interrupted by high clouds at 22:30 UT, 35 min after the event. A portion of data around the fourth contact was lost due to passing clouds.

An additional light curve, also occasionally affected by clouds, was acquired with the 1.52 m Cassini Telescope on 2015 August 25. The BFOSC detector was windowed to $8'.1 \times 4'.3$. Observations started at 20:30 August 25 UT, 130 min before the expected transit. They were ended 113 min after the event at 02:25 August 26 UT. Exposures of 120 s length were acquired through a Cousins $R$ filter.

\subsection{WASP-28 b}

A complete transit on 2010 September 17 was observed with the 0.4-m Schmidt-Cassegrain Meade LX200 GPS telescope at the Ankara University Observatory (Turkey).  An Apogee ALTA-U47 CCD camera with $1024 \times 1024$ 13-$\mu$m size pixels was used as detector. The instrument covered $11'  \times 11'$ on the sky. The time series were acquired between 18:49 September 17 UT and 01:28 September 18 UT, including 112 min and 81 min of monitoring before and after the transit. Fifteen-second long exposures were acquired through a Johnson $R$ filter.

A partial transit light curve was obtained on 2014 September 26. We used the 0.4 m Horacio Ghielmetti Telescope (THG) located at the Complejo Astron\'omico El Leoncito (Argentina), coupled with an Apogee Alta U16 CCD detector with a 4096 x 4096 matrix of 9 $\mu$m pixels. The FoV of the instrument was $31'  \times 31'$. Observations began at 02:30 UT, i.e. after the ingress phase due to some technical problems with the telescope. Observations were ended at 06:10 UT, 70 min after the egress. Images were acquired with the exposure time of 50 s through the Johnson $R$ filter.

The next complete transit light curve was obtained on 2014 October 02 with the 1.2-m Cassegrain telescope located at the Michael Adrian Observatory in Trebur (Germany). An SBIG STL-6303E CCD camera with $1536\times1024$ 18.0 $\mu$m pixels was used to record 35 s long exposures. The instrumental setup gave a $10'.0 \times 6'.7$ FoV. No filter was used. Observations began at 20:04 October 2 UT and ended at 01:44 October 3 UT. The host star was monitored 64 min before and 70 min after the transit. The telescope was manually guided during the run. Observing conditions were not photometric, and some of the measurements were affected by high humidity and varying atmospheric transparency.

\subsection{WASP-36 b}

A precise light curve of a complete transit was obtained on 2012 March 03 with the 2.0 m telescope at the National Astronomical Observatory Rozhen with the instrumental setup described in Section 2.1. Images were recorded without windowing that resulted in a circular FoV with a diameter of $16'$. Observations started at 19:31 UT, 25 min before the predicted transit beginning, and ended at 22:14 UT, 27 min after the transit end. Individual exposures, acquired through the Cousins $R$ filter, were 30 s long. The optics were slightly defocused, resulting in donut-like stellar profiles with diameters of $11''$. Manual guiding assured tracking inaccuracies smaller than $7''$.

\subsection{WASP-39 b}

A time series on 2012 February 26 was obtained with the 2.2 m Ritchey-Chr\'etien-Coud\'e telescope at the CAO. The Calar Alto Faint Object Spectrograph (CAFOS), equipped with a SITe\#1d matrix of $2048 \times 2048$ 24 $\mu$m pixels, was used in imaging mode. This multi-purpose instrument offers a FoV with a diameter of $16'$, but during our observations the field was windowed to $7'.3 \times 8'.9$  and $2 \times 2$ binning was applied. Observations began at 02:50 UT, 25 min before ingress, and were interrupted in the middle of the egress at 06:00 UT because of dawn. The telescope was significantly defocused to allow for 30 s long exposures in the Cousins $R$ filter. Donut-like stellar profiles had diameters of $28''$. The auto-guiding system kept the telescope pointed at the target star with a precision better than $1''$.

A transit on 2012 April 22 was observed with the 2.0 m telescope at Rozhen with the instrumental setup described in Section 2.1. We used the whole FoV of the instrument. Thanks to defocusing and $25''$ wide donut-like stellar profiles, 30 s long exposures in the Cousins $R$ filter were feasible. Monitoring began at 21:16 April 22 UT, 35 min before the transit, and ended 33 min after the event at 01:15 April 23 UT. Manual guiding guaranteed a tracking accuracy within $2''$. Observations were occasionally affected by passing high clouds.


\section{Literature datasets}

In addition to our new data, we also used the photometric and RV datasets, which are available in the literature. We give a short description of those datasets for the individual systems below.

\subsection{HAT-P-30 b}

There have been three high quality light curves published so far. Two time series come from Johnson \etal (2011) and were acquired with the 1.2 m telescope at the Fred Lawrence Whipple Observatory (FLWO) in Arizona (USA) on 2010 April 3 and November 22. A high-cadence time series obtained with the 2.0 m Liverpool Telescope at the Observatorio del Roque de los Muchachos on La Palma (Spain) on 2011 January 12 are provided by Enoch \etal (2011).

We used a set of Doppler measurements acquired with 4 instruments. Seventeen data points, acquired with the High-Resolution Echelle Spectrometer (HIRES; Vogt \etal 1994) on the Keck I telescope in Hawaii (USA) between 2010 April 27 and 2012 December 4, were taken from Knutson \etal (2014). This extended dataset includes reanalysed RV observations reported in the discovery paper by  Johnson \etal (2011). Johnson \etal (2011) also provide 11 RV measurements obtained between 2010 May 22 and 24 with the High-Dispersion Spectrograph (HDS; Noguchi \etal 2002) on the Subaru telescope in Hawaii. Enoch \etal (2011) present 10 RV observations acquired between 2010 October 10 and November 27 with the SOPHIE spectrograph (Bouchy \etal 2009), which is attached to the 1.93 m telescope at the Observatoire de Haute-Provence (France), and additional 10 measurements obtained between 2010 December 7 and 2011 January 4 with the CORALIE spectrograph mounted on the Swiss 1.2 m Leonhard Euler Telescope, which is located at the Geneva Observatory, a part of the La Silla Observatory (Chile).

\subsection{HAT-P-37 b}

The only three precise light curves which have been published so far come from the discovery paper by Bakos \etal (2012). They were acquired with the 1.2 m telescope at FLWO between February and April 2011. Only one of them, obtained on 2011 April 6, is complete. In addition, 13 RV measurements from Bakos \etal (2012) were used. Those RV measurements were acquired with HIRES between 2011 March 26 and May 17.

\subsection{TrES-5 b}

We used a set of 5 follow-up light curves acquired for 4 epochs (one transit was observed simultaneously in 2 bands) by Mandushev \etal (2011) and 9 light curves for 6 epochs (one transit was observed in 3 bands and the other one with two instruments) by Mislis \etal (2015). The first subset was acquired with the Lowell Observatory's 0.8 m, 1.1 m Hall, and 1.8 m Perkins telescopes between 2009 November 17 and 2010 November 30. The second one comes from the Cassini telescope, the 1.2 m and 2.2 m telescopes at CAO, and the 2.5 m Isaac Newton Telescope at La Palma (Spain). The subset was acquired between 2011 August 26 and 2013 September 14. Eight RV measurements were taken from Mandushev \etal (2011). They were derived from spectra acquired between 2010 September 17 and 2011 April 22 with the Tillinghast Reflector Echelle Spectrograph (TRES), mounted at the 60 inch Tillinghast Reflector at FLWO.

\subsection{WASP-28 b}

Anderson \etal (2015) present two precise transit light curves. The first dataset is incomplete and covers the egress phase only. It was acquired through a $z'$ filter with the 2 m Faulkes Telescope North (FTN) at the Haleakala Observatory in Hawaii on 2009 October 22. The complete light curve was obtained in a Gunn-$r$ filter with the Leonhard Euler Telescope on 2010 September 4. Petrucci \etal (2015) performed the first photometric follow-up observations with THG. Three complete transit light curves without any filter and one incomplete in the $R$ band were acquired on 2011 August 28, 2012 July 27, 2013 August 6, and October 27.

The RV data were taken from the discovery paper by Anderson \etal (2015). Twenty six measurements were acquired between 2009 June 24 and  2012 December 12 with the CORALIE spectrograph. An additional three were obtained on 2010 August 19 and 20 with the High Accuracy Radial Velocity Planet Searcher (HARPS) mounted on the 3.6 m ESO Telescope at the La Silla Observatory (Chile).

\subsection{WASP-36 b}

Eight light curves, acquired between 2010 December and 2011 January, are presented in the discovery paper by Smith \etal (2012). A transit on 2010 December 13 was observed simultaneously with two instruments, both located at La Silla Observatory: the Leonhard Euler Telescope with a Gunn-$r$ filter, and the 0.6 m robotic Transiting Planets and Planetesimals Small Telescope (TRAPPIST) without any filter. Both light curves are, however, partial. A complete light curve in the $z'$ band was acquired on 2010 December 25 with the FTN. In addition, one complete and two partial transits were observed without filters with the TRAPPIST on 2011 January 2, 5, and 8. A complete light curve from 2011 January 16 was acquired in a wide $V+R$ band with the Liverpool Telescope. The Leonhard Euler Telescope was again used to observe a full transit on 2011 January 22 through the Gunn-$r$ filter.

Smith \etal (2012) provide the only set of precise Doppler measurements for the WASP-36 star. Nineteen Doppler measurements were obtained with the CORALIE spectrograph between 2010 March 11 and 2011 January 11.

\subsection{WASP-39 b}

We used 2 light curves published in Faedi \etal (2011) that were acquired on 2010 May 18 and July 9. The first of them is complete and comes from the FTN. It was obtained using a Pan-STARRS $z$ filter. The second time series is partial and covers the ingress and a portion of the flat bottom phase of the transit. It was acquired with a Gunn $r$ filter with the Swiss 1.2-m Leonhard Euler Telescope. Ricci \etal (2015) provide photometric observations for two successive transits on 2014 March 17 and 21. The first of them was observed simultaneously in $U$ and $I$ bands with two separate instruments - the 2.12 m and 0.84 m Ritchey-Chr\'etien telescopes, respectively, at the San Pedro M\'artir National Astronomical Observatory (Mexico). Observations of the second transit were performed with the 0.84 m telescope through an $R$ filter.

We used Doppler measurements available in Faedi \etal (2011). Eight data points were acquired with the SOPHIE spectrograph in the high efficiency mode between 2010 April 8 and June 11. Additional RV observations were performed with the CORALIE spectrograph in 9 epochs between 2010 May 5 and July 3.


\section{Data reduction and analysis}

A complete data reduction process was performed with the AstroImageJ package\footnote{http://www.astro.louisville.edu/software/astroimagej} ({\sc AIJ}, Collins \etal 2016). Sequences of science images were corrected with dark or bias and flat-field frames. The mid-exposure dates were converted to barycentric Julian dates in barycentric dynamical time ($\rm{BJD_{TDB}}$). The light curves were obtained with the method of differential aperture photometry. The aperture radius was allowed to vary with a value depending on the full width at half maximum of the stellar profiles multiplied by a factor of between 0.8 and 1.4. The optimal value, giving the smallest scatter of the datapoints (typically 1.1), was determined for each dataset using trial runs. The sets of comparison stars were optimized to achieve the highest precision for the final light curves.

The light curves were detrended against airmass and position on the matrix if the data were acquired without telescope guiding. Some datasets were also detrended against the time domain. The detrending algorithm included in {\sc AIJ} offers the option of using not only data points acquired out of the transit but also in-transit data with a trial transit model. This feature is especially useful when the light curve covers only a portion of the transit. Despite giving timing uncertainties usually 1.5--2 times greater, incomplete transit light curves may still be useful for timing purposes. In such cases, we used the trial transit model with fixed parameters, the values of which were taken from modelling of complete light curves. The final fluxes were normalized to unity for out-of-transit brightness. The individual light curves are available online\footnote{http://www.home.umk.pl/\~{}gmac/TTV}.

For each transiting planet, our new light curves and those available in the literature were modelled simultaneously with the Transit Analysis Package ({\sc TAP}, Gazak \etal 2012). The transit model, which is based on the equations of  Mandel \& Agol (2002), is characterized by the orbital inclination $i_{b}$, the semimajor-axis scaled by the stellar radius $a_{b}/R_{*}$, the planetary to stellar radii ratio $R_{b}/R_{*}$, the mid-transit time $T_{mid}$, and limb-darkening (LD) coefficients of the quadratic law. Best-fitting parameters were found with the  Markov Chain Monte Carlo (MCMC) method, employing the Metropolis-Hastings algorithm and a Gibbs sampler. The main transit parameters, \ie $i_{b}$, $a_{b}/R_{*}$, and $R_{b}/R_{*}$, were linked together for all the light curves available for a given planetary system. Mid-transit times were linked for light curves of the same transit observed from two different sites or observed simultaneously in different filters. The values of the LD coefficients were allowed to vary under the Gaussian penalty of $\sigma = 0.1$ around theoretical values, which were linearly interpolated from tables of Claret \& Bloemen (2011) with an on-line tool\footnote{http://astroutils.astronomy.ohio-state.edu/exofast/limbdark.shtml} of the \textsc{EXOFAST} applet  (Eastman \etal 2013). In addition, we also allowed  fitting any possible linear trends in individual light curves to include their uncertainties in the total error budget of the fit.

We used 10 MCMC chains with $10^6$ steps each. To minimize the influence of the initial values of the parameters, the first 10\% of the results were discarded from each chain. Applying the wavelet-based technique of Carter \& Winn (2009) assures that the time-correlated (red) noise is included in the uncertainty estimates. To account for any possible asymmetry in the marginalized posterior probability distributions, a two-part Gaussian function was fitted. The mode of the fit was taken as the best-fitting value of the parameter, and left- and right-hand-side standard deviations were taken as its lower and upper 1$\sigma$ uncertainties.

New mid-transit times, together with those redetermined from literature data, were used to refine the transit  ephemeris for each transiting planet. The orbital period $P_b$ and the time of the reference epoch for cycle zero $T_0$ were determined in the result of a least-square linear fit, in which timing uncertainties were taken as weights. Errors of the fitted parameters were taken from the covariance matrix of the fit. The redetermined $P_b$ was used in the final runs of the light curve modelling with TAP.

To refine the semi-amplitude of stellar motion induced by a planetary companion $k_{\rm{b}}$, we used the Systemic 2.18 software (Meschiari \etal 2009). It allows one to simultaneously analyse RV and timing datasets. The implemented Levenberg-Marquardt algorithm was used to find the best-fitting values. The uncertainties were estimated with the bootstrap method based on $10^5$ trials. In initial iterations, we considered non-circular orbits with the eccentricity kept as a free parameter. We noticed however, that for all considered planets eccentricities were found to be consistent with zero within 1$\sigma$ uncertainties. Therefore, we adopted circular orbits in the final modelling and further calculations.


\section{Results}

For each system studied, the basic parameters from the RV and light curve modelling allowed us to determine the planetary mass $M_b$ (stellar masses and their uncertainties were taken from the discovery papers for the individual systems), radius $R_b$, mean density $\rho_b$, surface gravitational acceleration $g_b$, transit parameter $b_b = (a_b/R_*) \cos i$, orbital semi-major axis $a_b$, stellar radius $R_*$, mean stellar density $\rho_*$, and stellar surface gravity acceleration $g_*$. The value of $g_{b}$ was calculated following Eq.\ (7) of Southworth (2008), with a simplification for circular orbits:
\begin{equation}
     g_{b} = \frac{2\pi}{P_{b}}\frac{a_{b}^2 k_{b}}{R_{b}^{2} \sin i_{b}}\, . \;
\end{equation}
The value of $\rho_{*}$ was calculated with the formula based on Kepler's third law
\begin{equation}
     \rho_{*} = \frac{3\pi}{G P_{\rm{b}}^2} \left(\frac{a_{\rm{b}}}{R_{*}}\right)^3\, , \;
\end{equation}
which was derived under the assumption that $M_b \ll M_*$.

The quantities are collected in Table~2. The mid-transit times determined for the individual epochs, including the redetermined values for literature datasets, are given in Table~3. The residuals from linear ephemerides for transit timing are plotted in Fig.~2. Below we discuss the results obtained for the individual systems.


\subsection{HAT-P-30 b}

Our new transit light curve has a photometric noise ratio of 550 ppm and is the best quality follow-up photometric time series acquired for the system so far. There are discrepancies in the values of the parameters $i$ and $a/R_*$ between Johnson \etal (2011) and Enoch \etal (2011) at the level of 2.0 and $1.7 \, \sigma$, respectively. Our results with $i=82.70^{\circ}\pm0.19^{\circ}$ and $a/R_*=6.77^{+0.13}_{-0.12}$ are closer to the values reported by Enoch \etal (2011). We also find the remaining parameters to be consistent within  $1 \, \sigma$ to those of Enoch \etal (2011). The dataset of mid-transit times was found to be both accurate and precise to refine the transit ephemeris. The orbital period is redetermined with an uncertainty of 0.1 s that is 4--7 times smaller than in Johnson \etal (2011) and Enoch \etal (2011).


\subsection{HAT-P-37 b}

We provide four new transit light curves, which extend the time covered by the observations up to four years from the discovery of the system. Our determinations are consistent with those reported in Bakos \etal (2012) within $1 \, \sigma$. The new transit observations allowed us to determine the orbital period with an uncertainty of 0.14 s, \ie 4 times better than in the discovery paper.


\subsection{TrES-5 b}

Results of the follow-up observations by Mislis \etal (2015) are consistent with the values reported in the discovery paper by Mandushev \etal (2011). Our results also agree with both previous studies within $1 \, \sigma$. We note, however, that we improve the accuracy of $i_{b}$, $a_{b}/R_{*}$, and $R_{b}/R_{*}$ by $6.2\%$, $24\%$, and $30\%$, respectively, as compared to Mislis \etal (2015). We determine the orbital period with a precision of 15 ms that is 4 times better than in previous studies.


\begin{landscape}
\MakeTable{lcccccc}{18cm}{Parameters of the studied systems.}
{\hline
Parameter &  HAT-P-30 & HAT-P-37 & TrES-5 & WASP-28 & WASP-36 & WASP-39 \\
\hline
\multicolumn{7}{l}{Planetary properties}\\
~~$R_{\rm{b}}/R_{*}$ & $0.1109^{+0.0016}_{-0.0014}$ & $0.1394^{+0.0017}_{-0.0020}$ & $0.14203^{+0.00084}_{0.00091}$  & $0.1160^{+0.0017}_{-0.0014}$  & $0.1391^{+0.0011}_{-0.0012}$ & $0.1457^{+0.0015}_{-0.0016}$ \\
~~$M_{\rm{b}}$ ($M_{\rm{Jup}}$) & $0.733\pm0.023$ & $1.19\pm0.11$ & $1.804\pm0.076$ & $0.927\pm0.049$ & $2.295\pm0.058$ & $0.283\pm0.041$\\
~~$R_{\rm{b}}$ ($R_{\rm{Jup}}$) & $1.469^{+0.039}_{-0.037}$ & $1.231^{+0.048}_{-0.043}$ & $1.203^{+0.021}_{-0.020}$ & $1.250^{+0.029}_{-0.053}$ & $1.330^{+0.030}_{-0.029}$ & $1.332^{+0.034}_{-0.031}$\\
~~$\rho_{\rm{b}}$ ($\rho_{\rm{Jup}}$) & $0.232^{+0.020}_{-0.019}$ & $0.638^{+0.095}_{-0.089}$ & $1.035^{+0.070}_{-0.068}$ & $0.474^{+0.041}_{-0.065}$ & $0.976^{+0.070}_{-0.068}$ & $0.120^{+0.020}_{-0.019}$ \\
~~$g_{\rm{b}}$ (m~s$^{-2}$) & $8.82^{+0.48}_{-0.45}$ & $20.4^{+2.3}_{-2.2}$ & $32.3^{+1.6}_{-1.5}$ & $15.4^{+0.8}_{-1.4}$ & $33.7^{+1.5}_{-1.4}$ & $4.14^{+0.62}_{-0.61}$\\
\multicolumn{7}{l}{Orbital properties}\\
~~$i_{\rm{b}}$ ($^{\circ}$) & $82.70\pm0.19$ & $86.67^{+0.40}_{-0.33}$ & $84.65^{+0.24}_{-0.22}$ & $88.35^{+1.65}_{-0.92}$ & $83.62^{+0.30}_{-0.26}$ & $87.75^{+0.27}_{-0.20}$ \\
~~$b_{\rm{b}}$ & $0.860^{+0.028}_{-0.27}$ & $0.533^{+0.055}_{-0.065}$ & $0.577^{+0.025}_{-0.027}$ &  $0.25^{+0.14}_{-0.25}$ & $0.657^{+0.029}_{-0.033}$ & $0.447^{+0.041}_{-0.055}$ \\
~~$k_{\rm{b}}$ (m~s$^{-1}$) & $90.7\pm2.1$ & $180\pm16$ & $345\pm13$ & $123.4\pm5.1$ & $391.4\pm6.0$ & $37.9\pm5.4$ \\
~~$a_{\rm{b}}$ (AU) & $0.04190\pm0.00046$ & $0.03791\pm0.00059$ & $0.02450\pm0.00022$ & $0.04464\pm0.00073$ & $0.02641\pm0.00026$ & $0.04858\pm0.00052$\\
~~$P_{\rm{b}}$ (d) & $2.8106084$ & $2.7974471$ & $1.48224754$ & $3.4088387$ & $1.5373639 $ & $4.0552765$\\
 & $\pm 0.0000011$ & $\pm 0.0000011$ & $\pm 0.00000017$ & $\pm 0.0000016$ & $\pm 0.0000014$ & $\pm 0.0000035$\\
~~$T_{\rm{0}}$ (BJD$_{\rm{TDB}}$ 2450000+) & $5456.46620$ & $5642.143820$ & $5443.25271$ & $5290.40595$ & $5569.83802$ & $5342.96982$\\
 & $\pm 0.00013$ & $\pm 0.000156$ & $\pm 0.00011$ & $\pm 0.00046$ & $\pm 0.00013$ & $\pm 0.00051$\\
\multicolumn{7}{l}{Stellar properties}\\
~~$a_{\rm{b}}/R_{*}$ & $6.771^{+0.013}_{-0.012}$ & $9.19^{+0.31}_{-0.25}$ & $6.188^{+0.085}_{-0.078}$ & $8.859^{+0.067}_{-0.326}$  & $5.91^{+0.11}_{-0.10}$ & $11.37^{+0.24}_{-0.20}$  \\
~~$R_{*}$  ($R_{\odot}$) & $1.330^{+0.030}_{-0.028}$ & $0.887^{+0.033}_{-0.028}$ & $0.851^{+0.014}_{-0.013}$ & $1.083^{+0.020}_{-0.044}$ & $0.960^{+0.020}_{-0.019}$ & $0.918^{+0.022}_{-0.019}$ \\
~~$\rho_{*}$  $(\rho_{\odot})$ & $0.528^{+0.031}_{-0.029}$ & $1.33^{+0.13}_{-0.11}$ & $1.449^{+0.060}_{-0.055}$ & $0.804^{+0.018}_{-0.089}$ & $1.176^{+0.066}_{-0.060}$ & $1.201^{+0.075}_{-0.063}$ \\
~~$\log g_{*}$ (cgs units) & $4.284^{+0.028}_{-0.026}$ & $4.510^{+0.046}_{-0.038}$ & $4.529^{+0.019}_{-0.018}$ & $4.378^{+0.013}_{-0.051}$ & $4.490^{+0.026}_{-0.024}$ & $4.480^{+0.029}_{-0.025}$ \\
\hline
}
\end{landscape}

\MakeTable{lrcll}{12.5cm}{New transit times for the studied planets. The value of O--C is the difference between observed and predicted mid-transit times, calculated according to the refined ephemerides.}
{\hline
Date UT & Epoch & $T_{\rm{mid}}$ (BJD$_{\rm{TDB}}$ 2450000+) & ~~O--C (d)  & Data source \\
\hline
\multicolumn{5}{c}{HAT-P-30 b}\\
 2010 Apr 4  & --59  & $5290.64029^{+0.00111}_{-0.00095}$ & $-0.00001$ & Johnson \etal (2011)\\
 2010 Nov 23  & 24  & $5523.92055^{+0.00083}_{-0.00081}$ & $-0.00025$ & Johnson \etal (2011)\\
 2011 Jan 13  & 42  & $5574.51188^{+0.00046}_{-0.00054}$ & $+0.00012$ & Enoch \etal (2011)\\
 2012 Jan 19  & 174  & $5945.51205^{+0.00054}_{-0.00049}$ & $-0.00002$ & This paper\\
\multicolumn{5}{c}{HAT-P-37 b}\\
 2011 Feb 24  & --9  & $5616.96715^{+0.00080}_{-0.00087}$ & $+0.00018$ & Bakos \etal (2012)\\
 2011 Mar 24  &   1  & $5644.94118^{+0.00038}_{-0.00038}$ & $-0.00024$ & Bakos \etal (2012)\\
 2011 Apr 07  & 6  & $5658.92864^{+0.00052}_{-0.00054}$ & $+0.00000$ & Bakos \etal (2012)\\
 2012 Jul 25  & 176  & $6134.49387^{+0.00076}_{-0.00080}$ & $-0.00028$ & This paper\\
 2012 Oct 03  & 201  & $6204.43087^{+0.00078}_{-0.00070}$ & $+0.00065$ & This paper\\
 2013 Sep 29  & 330  & $6565.30156^{+0.00075}_{-0.00074}$ & $+0.00103$ & This paper\\
 2015 Sep 29  & 591  & $7295.4317^{+0.0012}_{-0.0011}$ & $-0.0018$ & This paper\\
\multicolumn{5}{c}{TrES-5 b}\\
 2009 Nov 17  & --196  & $5152.73184^{+0.00093}_{-0.00080}$ & $-0.00035$ & Mandushev \etal (2011)\\
 2010 Jun 05  & --61  & $5352.83535	^{+0.00029}_{-0.00028}$ & $-0.00026$ & Mandushev \etal (2011)\\
 2010 Sep 05  & 1  & $5444.73500^{+0.00052}_{-0.00053}$ & $+0.00005$ & Mandushev \etal (2011)\\
 2010 Nov 30  & 59  & $5530.70451^{+0.00056}_{-0.00046}$ & $-0.00081$ & Mandushev \etal (2011)\\
 2011 Aug 26  & 241  & $5800.47470^{+0.00024}_{-0.00022}$ & $+0.00033$ & Mislis \etal (2015)\\
 2011 Sep 04  & 247  & $5809.36785^{+0.00027}_{-0.00027}$ & $+0.00000$ & Mislis \etal (2015)\\
 2012 Sep 10  & 498  & $6181.41212^{+0.00029}_{-0.00028}$ & $+0.00013$ & Mislis \etal (2015)\\
 2013 Jun 15  & 685  & $6458.59213^{+0.00048}_{-0.00049}$ & $-0.00014$ & Mislis \etal (2015)\\
 2013 Jul 31  & 716  & $6504.54182^{+0.00029}_{-0.00030}$ & $-0.00013$ & Mislis \etal (2015)\\
 2013 Sep 14  & 747  & $6550.49157^{+0.00021}_{-0.00020}$ & $-0.00005$ & Mislis \etal (2015)\\
 2013 Nov 08  & 784  & $6605.33486^{+0.00023}_{-0.00021}$ & $+0.00008$ & This paper\\
 2014 Aug 21  & 977  & $6891.40861^{+0.00044}_{-0.00043}$ & $+0.00005$ & This paper\\
 2015 Aug 25  & 1226  & $7260.48794^{+0.00034}_{-0.00032}$ & $-0.00026$ & This paper\\
 2015 Sep 03  & 1232  & $7269.38183^{+0.00052}_{-0.00053}$ & $+0.00015$ & This paper\\
\multicolumn{5}{c}{WASP-28 b}\\
 2009 Oct 22  &  --48  & $5126.7817^{+0.0011}_{-0.0011}$ & $+0.0000$ & Anderson \etal (2015)\\
 2010 Sep 04  &  45  & $5443.80414^{+0.00061}_{-0.00063}$ & $+0.00045$ & Anderson \etal (2015)\\
 2010 Sep 17  &   49  & $5457.4375^{+0.0013}_{-0.0014}$ & $-0.0015$ & This paper\\
 2011 Aug 28  &  150  & $5801.7320^{+0.0018}_{-0.0017}$ & $+0.0003$ & Petrucci \etal (2015)\\
 2012 Jul 27  &  248  & $6135.7958^{+0.0016}_{-0.0016}$ & $-0.0021$ & Petrucci \etal (2015)\\
 2013 Aug 06  &  358  & $6510.7618^{+0.0022}_{-0.0020}$ & $-0.0084$ & Petrucci \etal (2015)\\
 2013 Oct 27  &  382  & $6592.5831^{+0.0017}_{-0.0015}$ & $+0.0008$ & Petrucci \etal (2015)\\
 2014 Sep 26  & 480  & $6926.6468^{+0.0021}_{-0.0010}$ & $-0.0017$ & This paper\\
 2014 Oct 02  & 482  & $6933.46649^{+0.00071}_{-0.00073}$ & $+0.00029$ & This paper\\
\multicolumn{5}{c}{WASP-36 b}\\
 2010 Dec 13  & --17  & $5543.70256^{+0.00047}_{-0.00045}$ & $-0.00027$ & Smith \etal (2012)\\
 2010 Dec 25  & --9  & $5556.00284^{+0.00058}_{-0.00059}$ & $+0.00109$ & Smith \etal (2012)\\
 2011 Jan 02  & --4  & $5563.68872^{+0.00033}_{-0.00031}$ & $+0.00016$ & Smith \etal (2012)\\
 2011 Jan 05  & --2  & $5566.76321^{+0.00055}_{-0.00056}$ & $-0.00009$ & Smith \etal (2012)\\
 2011 Jan 08  & 0  & $5569.83778^{+0.00038}_{-0.00036}$ & $-0.00024$ & Smith \etal (2012)\\
 2011 Jan 16  & 5  & $5577.52480^{+0.00040}_{-0.00039}$ & $-0.00004$ & Smith \etal (2012)\\
 2011 Jan 22  & 9  & $5583.67420^{+0.00026}_{-0.00024}$ & $-0.00010$ & Smith \etal (2012)\\
 2012 Mar 16  & 282  & $6003.37465^{+0.00044}_{-0.00041}$ & $+0.00002$ & This paper\\
\multicolumn{5}{c}{WASP-39 b}\\
 2010 May 18  & --2  & $5334.85919^{+0.00058}_{-0.00055}$ & $-0.00008$ & Faedi \etal (2011)\\
 2012 Jul 10  & 11  & $5387.57818^{+0.00066}_{-0.00078}$ & $+0.00032$ & Faedi \etal (2011)\\
 2012 Feb 26  & 158  & $5983.70314^{+0.00035}_{-0.00035}$ & $-0.00038$ & This paper\\
 2012 Apr 22  & 172  & $6040.47773^{+0.00038}_{-0.00035}$ & $+0.00035$ & This paper\\
 2014 Mar 17  & 343 & $6733.93612^{+0.00048}_{-0.00043}$ & $+0.00645$ & Ricci \etal (2015)\\
 2014 Mar 21  & 344 & $6737.99199^{+0.00043}_{-0.00041}$ & $+0.00704$ & Ricci \etal (2015)\\
 \hline
}

\begin{figure}[thb]
\begin{center}
\includegraphics[width=1.0\textwidth]{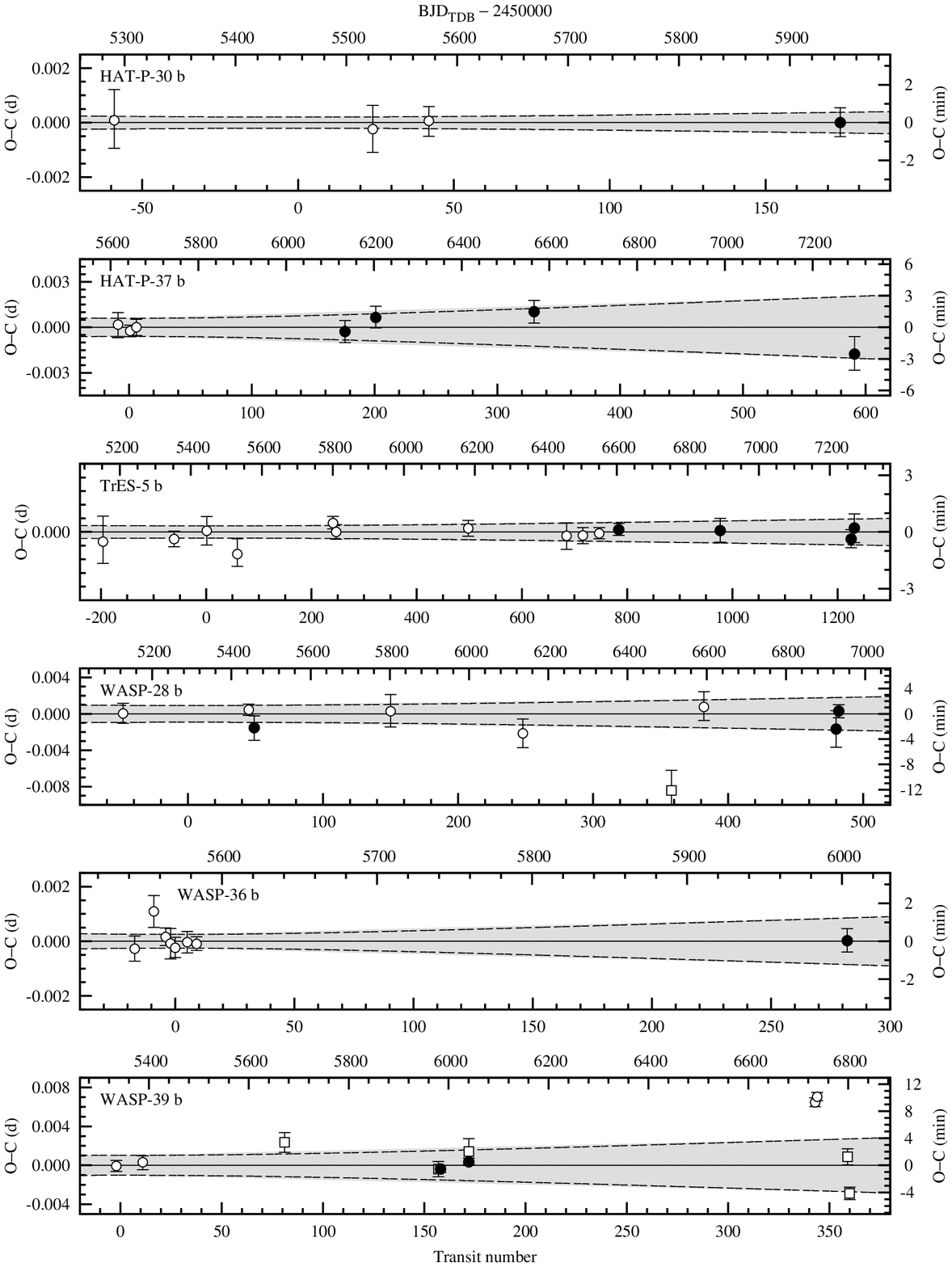}
\end{center}
\FigCap{Residuals from linear ephemerides for transit timing of the studied systems. Results of our new observations are marked with filled circles. Open symbols denotes mid-transit times redetermined from literature data (see text for details). The grey areas bounded by dashed lines show the propagation of the ephemeris uncertainties at a 95.5\% confidence level.}
\end{figure}


\subsection{WASP-28 b}

We provide three complete transit light curves, one of which is the first high-precision photometric time series of the transit of WASP-28~b, acquired 482 epochs from the cycle zero defined by the ephemeris given by Anderson \etal (2015). The redetermined system parameters agree with previous values within the $1 \, \sigma$ uncertainties. The analysis of timing residuals from the linear ephemeris shows that the mid-transit time derived from the incomplete light curve of Petrucci \etal (2015) is a 4$\sigma$ outlier. It is marked with an open square symbol in the timing residuals plot in Fig.~2. We skipped this point in the final calculations of the ephemeris. The orbital period is refined with an error of 0.14 s that is four times smaller than in Anderson \etal (2015) and 2 times smaller than in Petrucci \etal (2015).


\subsection{WASP-36 b}

Our high precision photometric time series, acquired 282 epochs after the initial ephemeris given in Smith \etal (2012), is the first follow-up transit observation for WASP-36~b. It was corrected for third light following the procedure used in Smith \etal (2012). The results of our modelling agree within $1 \, \sigma$ with the values determined by Smith \etal (2012). The new precise mid-transit time allowed us to refine the orbital period and to reduce its uncertainty by a factor of 2.6 compared to Smith \etal (2012).


\subsection{WASP-39 b}

In trial iterations, we considered two new transit light curves and all 5 photometric datasets from the literature. We noticed, however, that the light curve obtained in the $U$-band by Ricci \etal (2015) exhibits significant deformations. They are particularly noticeable in the second part of the transit with observed points systematically above the fit. Those deformations do not correspond to any features in the $I$-band light curve, so are probably of instrumental origin or caused by variable weather conditions, which were reported to be non-photometric. A trial fit without the $U$-band light curve resulted in smaller uncertainties in $R_p/R_*$ and $a/R_*$. We also noted that including this light curve produces little improvement in the transit time. So, in the final fit, we skipped the $U$-band dataset and limited the sample to 7 light curves obtained in red filters only.

Refined system parameters are in agreement with previous values reported by Faedi \etal (2011) and Ricci \etal (2015). The only exception is the orbital period, which is reported by Ricci \etal (2015) to be longer by about 3 s than the value of Faedi \etal (2011). Our result $P_b=4.0552766 \pm 0.0000035$ d is consistent within 1.4 $\sigma$ with  $P_b=4.055259 \pm 0.000009$ d derived by Faedi \etal (2011), and differs by 5.0 $\sigma$ from the value of Ricci \etal (2015). We used additional transit times from amateur observations that are publicly available in the Exoplanet Transit Database (Poddany \etal 2010). We selected six of the most reliable determinations which are based on transit light curves with the quality mark of 3 or better. The transit at epoch 172 was observed with two separate instruments, so both individual mid-transit times and their uncertainties were averaged. As is seen in the diagram of the transit time residuals from the linear ephemeris  in Fig.~2, these additional data, marked with open square symbols, are consistent with our ephemeris. Mid-transit times of Ricci \etal (2015) in epochs 343 and 344 stand out by 9.7 minutes. The reason of this systematic shift in time stamps remains unexplained\footnote{We suspect that Barycentric time correction could be applied twice by Ricci \etal (2015) but we have not received any confirmation from the authors.}.


\section{Concluding discussion}

We present 15 new transit light curves for 6 transiting exoplanets: HAT-P-30~b, HAT-P-37~b, TrES-5~b, WASP-28~b, WASP-36~b, and WASP-39~b. For three systems, this is the first follow-up study, and for another two we bring the first high-precision transit light curves. Our new data were combined with photometric and RV time series, which are available in the literature, in order to redetermine the system parameters in a homogeneous way. For all the investigated systems, our determinations are in good agreement with the results of previous studies. New transit times allowed us to refine transit ephemerides and reduce uncertainties for orbital periods by a factor between 2 and 7. We found that the observed transits follow linear ephemerides, and give no hint of variations in transit times.


\Acknow{We are grateful to Dr.~Susana Cristina Cabral de Barros, Dr.~Georgi Mandushev, and Dr.~Alexis Smith for sharing the HAT-P-30, TrES-5, and WASP-36 light curves with us. We would like to thank the Calar Alto staff for their help during observing runs. GD acknowledges Regione Campania for support from POR-FSE Campania 2014-2020. Part of this paper is the result of the exchange and joint research project {\em Spectral and photometric studies of variable stars} between the Polish and Bulgarian Academies of Sciences. This research has made use of the Exoplanet Transit Database, maintained by the Variable Star Section of the Czech Astronomical Society.}

\end{document}